\documentclass{IEEEtran}
\usepackage{cite}
\usepackage{amsmath,amssymb,amsfonts}
\usepackage{algorithmic}
\usepackage{latexsym}
\usepackage{graphicx}
\usepackage{textcomp}
\def\BibTeX{{\rm B\kern-.05em{\sc i\kern-.025em b}\kern-.08em
		T\kern-.1667em\lower.7ex\hbox{E}\kern-.125emX}}

\DeclareGraphicsExtensions{.pdf,.jpeg,.png}
\usepackage{cite}
\usepackage{url}
\usepackage{soul}
\usepackage{color}
\usepackage{booktabs}
\usepackage{multirow}
\usepackage[normalem]{ulem}
\usepackage{xcolor}

\newcommand{\BE}{\begin{equation}}   
	\newcommand{\EE}{\end{equation}}     
\newcommand{\BEn}{\begin{equation*}} 
	\newcommand{\EEn}{\end{equation*}}   
\newcommand{\BF}{\begin{figure}}     
	\newcommand{\EF}{\end{figure}}  	 
\newcommand{\BFw}{\begin{figure*}}   
	\newcommand{\EFw}{\end{figure*}}     
\newcommand{\BT}{\begin{table}}		 
	\newcommand{\ET}{\end{table}}        

\newcommand{\J}{\mathrm{j}}                 

\providecommand{\D}{\,\mathrm{d}}           
\providecommand{\V}[1]{\boldsymbol{#1}}     
\newcommand{\UV}[1]{\hat{\V{#1}}}               

\providecommand{\ZVAC}{Z_0}  
\providecommand{\Zsurf}{R_\mathrm{s}}

\newcommand{\CAP}{\mathrm{e}} 
\newcommand{\IND}{\mathrm{m}} 

\providecommand{\Prad}{P_\mathrm{rad}}
\providecommand{\Preact}{P_\mathrm{react}}
\providecommand{\Plost}{P_\mathrm{lost}}
\providecommand{\Jcap}{\V{J}^\CAP}
\providecommand{\Jind}{\V{J}^\IND}
\providecommand{\betacap}{\beta^\CAP}
\providecommand{\betaind}{\beta^\IND}
\providecommand{\Preactcap}{\Preact^\CAP}
\providecommand{\Preactind}{\Preact^\IND}
\providecommand{\Plostcap}{\Plost^\CAP}
\providecommand{\Plostind}{\Plost^\IND}
\providecommand{\Pradcap}{\Prad^\CAP}
\providecommand{\Pradind}{\Prad^\IND}

\usepackage[colorinlistoftodos,prependcaption,textsize=small]{todonotes}
\usepackage{regexpatch}
\makeatletter
\xpatchcmd{\@todo}{\setkeys{todonotes}{#1}}{\setkeys{todonotes}{inline,#1}}{}{}
\makeatother

\colorlet{colorMG}{red!30!black}

\begin{document}
	
	\title{{\fontsize{24}{26}\selectfont{Communication\rule{24pc}{0.5pt}}}\break\fontsize{16}{18}\selectfont
		Dissipation Factors of Spherical Current Modes on Multiple Spherical Layers}
	\author{Vit~Losenicky, Lukas~Jelinek, Miloslav~Capek,~\IEEEmembership{Senior Member,~IEEE} and Mats~Gustafsson~\IEEEmembership{Senior Member,~IEEE}
		\thanks{Manuscript received XXX, 2018; revised XXX, 2018.}
		\thanks{This work was supported by  the  Grant  Agency  of  the  Czech  Technical  University in  Prague  under Grant SGS16/226/OHK3/3T/13.}
		\thanks{V.~Losenicky, L.~Jelinek and M.~Capek are with the Department of Electromagnetic Field, Faculty of Electrical Engineering, Czech Technical University in Prague, Technicka~2, 16627, Prague, Czech Republic
			(e-mail: \mbox{losenvit@fel.cvut.cz}, \mbox{lukas.jelinek@fel.cvut.cz}, \mbox{miloslav.capek@fel.cvut.cz}).}%
		\thanks{M. Gustafsson is with the Department of Electrical and Information Technology, Lund University, Box 118, SE-221 00 Lund, Sweden. (email: mats.gustafsson@eit.lth.se).}%
	}
	
	
	\maketitle
	
	\begin{abstract}
		Radiation efficiencies of modal current densities distributed on a spherical shell are evaluated in terms of dissipation factor. The presented approach is rigorous, yet simple and straightforward, leading to closed-form expressions. The same approach is utilized for a two-layered shell and the results are compared with other models existing in the literature. Discrepancies in this comparison are reported and reasons are analyzed. Finally, it is demonstrated that radiation efficiency potentially benefits from the use of internal volume which contrasts with the case of the radiation Q-factor.
	\end{abstract}
	
	
	\section{Introduction}
	\label{intro}
	
	\IEEEPARstart{T}{he} fundamental bounds on radiation efficiency have become increasingly interesting in recent years \cite{Pfeiffer_FundamentalEfficiencyLimtisForESA,Thal2018_RadiationEfficiencyLimits,Jelinek+etal2017} as low radiation efficiency, together with a high radiation Q-factor presents a serious performance bottleneck for all electrically small antenna designs~\cite{VolakisChenFujimoto_SmallAntennas}.
	
	Similar to fundamental bounds on radiation Q-factor, fundamental bounds on radiation efficiency were first approached using the example of a spherical shell. The reason is twofold. First, the mathematics of spherical modes is analytically tractable. Second, it has been assumed \cite{Pfeiffer_FundamentalEfficiencyLimtisForESA} that, analogous to radiation Q-factor, the best radiation efficiency belongs to surface spherical modes.
	
	The major purpose of this communication is to extend the study presented in \cite{Pfeiffer_FundamentalEfficiencyLimtisForESA} by providing a full-wave treatment of multilayer scenarios and to provide evidence that the surface spherical currents do not form a lower bound to the dissipation factor of a general volumetric radiator. Considering the practical demand on resonance and the fact that loss-less external tuning is unreachable \cite{Smith_1977_TAP,Pfeiffer_FundamentalEfficiencyLimtisForESA,Thal2018_RadiationEfficiencyLimits,Jelinek+etal2017}, here, attention is primarily paid to self-resonant current densities, and the externally tuned results are considered only as intermediate products. All analytical results are verified with full-wave numerical calculations.
	
	{The communication is organized as follows. In Section~\ref{Sec:OneShell}, the lowest dissipation factor for a single spherical shell is derived and compared with existing results. The model is generalized to two spherical layers in Section~\ref{Sec:TwoShells}, and to multiple layers in Section~\ref{Sec:NShells}. The communication is concluded in Section~\ref{Concl}.
		
		\section{Dissipation factor of a single spherical layer}
		\label{Sec:OneShell}
		
		This section reformulates the results presented in \cite{Pfeiffer_FundamentalEfficiencyLimtisForESA} by directly manipulating vector spherical waves~\cite{Kristensson_ScatteringBook}. Some discrepancies in the model used in \cite{Pfeiffer_FundamentalEfficiencyLimtisForESA} are also indicated.
		
		It is possible to show~\cite{Stratton_ElectromagneticTheory} that, within a time-harmonic steady state, electric field~$\V{E}$ and surface current density~$\V{J}$, corresponding to the modes of a spherical layer of radius~$a$, read
		\begin{align}
			\label{SpherModes1}
			\V{E}_{mn}^\mathrm{TE} & = -\ZVAC \, \zeta_n \left( ka \right) \psi'_n \left( ka \right) \V{M}_{mn}, \\
			\label{SpherModes2}
			\V{E}_{mn}^\mathrm{TM} &= \ZVAC \, \psi_n \left( ka \right) \zeta'_n \left( ka \right) \V{N}_{mn}, \\
			\label{SpherModes3}
			\V{J}_{mn}^\mathrm{TE} &= \UV{r} \times \V{N}_{mn}, \\
			\label{SpherModes4}
			\V{J}_{mn}^\mathrm{TM} &= \UV{r} \times \V{M}_{mn},
		\end{align}  
		where
		\begin{align}
			\label{SpherModes5}
			\psi_n \left( x \right) &= x \mathrm{j}_n \left( x \right), \\
			\chi_n \left( x \right) &= - x \mathrm{y}_n \left( x \right), \\
			\zeta_n \left( x \right) &= x \mathrm{h}_n^{\left( 2 \right)} \left( x \right) = \psi_n \left( x \right) + \J \chi_n \left( x \right),
		\end{align}
		are Riccati-Bessel functions, the symbol $'$ denotes differentiation, $\ZVAC$ is the free-space impedance, $k$ is the free-space wavenumber, \mbox{$\mathrm{j}_n$}, \mbox{$\mathrm{y}_n$ and \mbox{$\mathrm{h}_n^{\left( 2 \right)}$}} are the spherical Bessel's functions of order~$n$~\cite{Arfken_MathForPhysicists}, functions $\V{M}$ and $\V{N}$ are spherical vector waves defined in~\cite{Stratton_ElectromagneticTheory} with Bessel's function \mbox{$\mathrm{j}_n$} inserted, and~$\UV{r}$ is the unit vector pointing in the radial direction. Electric field $\V{E}_{mn}^\mathrm{TE/TM}$ and current density $\V{J}_{mn}^\mathrm{TE/TM}$ also depend on spherical angular variables, but this dependence is of no relevance in this paper. 
		
		In order to evaluate radiation efficiency $\eta$ of modal current distributions, this paper uses dissipation factor $\delta$ \cite{Harrington_AntennaExcitationForMaximumGain} defined via $\eta = 1/ \left( 1+\delta\right)$. Dissipation factor is thus the ratio of the cycle mean power lost by conduction and cycle mean power lost by radiation.
		
		In order to evaluate dissipation factors of surface current distributions, the complex power \cite{Harrington_TimeHarmonicElmagField}
		\BE
		\label{Pcomplex}
		\Prad +\J \Preact = - \frac{1}{2} \int\limits_S \V{J}^\ast \cdot \V{E} \D{S}
		\EE
		and cycle mean lost power
		\BE
		\label{Plost}
		\Plost = \frac{\Zsurf}{2} \int\limits_S \V{J}^\ast \cdot \V{J} \D{S}
		\EE
		are needed, where $^\ast$ denotes complex conjugation, and $\Zsurf$ denotes surface resistance (homogeneously distributed over the surface $S$). For current densities flowing on highly conducting bodies, a surface resistance model \mbox{$\Zsurf=1/(\sigma d)$} can be assumed, with $d$ being an effective penetration distance of the field into the conductor. For electrically thick conductors, $d$ can be put equal to the penetration depth~\cite{Jackson_ClassicalElectrodynamics}.
		\\
		\\
		In line with \cite{Pfeiffer_FundamentalEfficiencyLimtisForESA} and considering a major cost of resonance tuning to radiation efficiency~\cite{Jelinek+etal2017} let us also prepare the grounds to form a resonant combination of selected spherical modes. To that point suppose a current density 
		\BE
		\label{modeCombination}
		\V{J} = \Jcap + \alpha \Jind
		\EE
		with tuning coefficient
		\BE
		\label{constant}
		\left| \alpha \right|^2 = - \displaystyle\frac{\Preactcap}{\Preactind}
		\EE
		is formed with $\Jcap$ and $\Jind$ being capacitative and inductive (excess electric or magnetic energy) spherical modes \eqref{SpherModes3}, \eqref{SpherModes4}, and $\Preactcap$, $\Preactind$ being the corresponding reactive powers~\eqref{Pcomplex}. Owing to the orthogonality of spherical modes~\cite{Stratton_ElectromagneticTheory}, the current density~\eqref{modeCombination} is self-resonant with~\mbox{$\Preact = 0$}. 
		
		The dissipation factor~$\delta$, corresponding to the current density \eqref{modeCombination}, reads
		\BE
		\label{dissipGeneral}
		\delta = \frac{\Plost}{\Prad} = \frac{\Plostcap + \left| \alpha \right|^2 \Plostind}{\Pradcap + \left| \alpha \right|^2 \Pradind} = \frac{ \delta^\CAP - \displaystyle\frac{\lambda^\CAP}{\lambda^\IND} \delta^\IND}{1 - \displaystyle\frac{\lambda^\CAP}{\lambda^\IND}},
		\EE
		where mode orthogonality has once more been employed and where normalized reactances
		\BE
		\label{lambda}
		\lambda^{\CAP/\IND} = \frac{\Preact^{\CAP/\IND}}{\Prad^{\CAP/\IND}}.
		\EE
		were defined.
		
		At small electrical sizes, where bounds on dissipation are of interest, the capacitive modes $^\CAP$ are the spherical $\mathrm{TM}$ modes, while inductive modes $^\IND$ are the spherical $\mathrm{TE}$ modes. The last step prior to evaluation of (\ref{dissipGeneral}) is thus to find the dissipation factors $\delta _n^\mathrm{TE/TM}$. The substitution of~(\ref{SpherModes1})--(\ref{SpherModes4}) into~(\ref{Pcomplex}) and~(\ref{Plost}) leads to 
		\begin{align}
			\label{modalDelta1}
			\delta_{\circ,n}^\mathrm{TE} &= \frac{\Zsurf}{\ZVAC} \frac{1}{ \left( \psi_n \left( ka \right) \right)^2}, \\
			\label{modalDelta2}
			\delta _{\circ,n}^\mathrm{TM} &= \frac{\Zsurf}{\ZVAC} \frac{1}{ \left( \psi'_n \left( ka \right) \right)^2},
		\end{align}
		and to the expressions for the normalized reactances
		\begin{align}
			\label{modalLambda1}
			\lambda _{\circ,n}^\mathrm{TE} &= \frac{\chi_n \left( ka \right)}{\psi_n \left( ka \right)}, \\
			\label{modalLambda2}
			\lambda _{\circ,n}^\mathrm{TM} &= \frac{\chi'_n \left( ka \right)}{ \psi'_n \left( ka \right)},
		\end{align}
		which are equal to the characteristic numbers of a perfectly conducting spherical layer~\cite{CapekEtAl_ValidatingCMsolvers}. Subindex $\circ$ in (\ref{modalDelta1})--(\ref{modalLambda2}) denotes quantities corresponding to a single spherical layer.
		
		The dissipation factor~\eqref{dissipGeneral} of any resonant combination of two spherical modes on a single spherical layer can easily be evaluated by substituting \eqref{modalDelta1}--\eqref{modalLambda2} into \eqref{dissipGeneral}. 
		
		A direct comparison with dissipation factors evaluated in~\cite{Pfeiffer_FundamentalEfficiencyLimtisForESA} reveals that the dissipation factors evaluated above are approximately two times higher. The reason for this discrepancy is the assumption\footnote{We would like to thank C.~Pfeiffer for pointing this out to us during a private discussion.} made in~\cite{Pfeiffer_FundamentalEfficiencyLimtisForESA} that the spherical shell is composed of an inner and outer surface, both exhibiting the same surface resistance~$\Zsurf$. Assuming that the radial distance between the layers is negligible with respect to wavelength, it is easy to prove that such a configuration leads exactly to two times lower dissipation factors when compared to a single layer\footnote{The two-layer scenario can be understood as a transformation $d\to 2d$ within the surface resistance model \mbox{$\Zsurf=1/(\sigma d)$} which leads to $\Zsurf \to \Zsurf / 2$ and thus to two times smaller dissipation factors according to (\ref{modalDelta1}) and (\ref{modalDelta2}).}. The reason is that the radiated power increases four times (due to cross terms in the $\V{E} \cdot \V{J}^\ast$ product), while losses increase only by a factor of two (having no cross terms in the $\V{J} \cdot \V{J}^\ast$ product). The mathematical proof is given in the next section. 
		
		Taking into account the above-mentioned factor of two (multiplying the results of~\cite{Pfeiffer_FundamentalEfficiencyLimtisForESA} by two), the comparison of results derived here and the results derived in~\cite{Pfeiffer_FundamentalEfficiencyLimtisForESA} is shown in Fig.~\ref{fig1} and, simultaneously, in Table~\ref{tab1}, adopting the naming convention from~\cite{Pfeiffer_FundamentalEfficiencyLimtisForESA}. The results presented here coincide with those derived in~\cite{Thal2018_RadiationEfficiencyLimits} and are well approximated by the results derived in~\cite{Pfeiffer_FundamentalEfficiencyLimtisForESA}. With respect to the comparison it is also important to note a considerable difference between asymptotic formulas and full wave results which, in the \mbox{$\mathrm{TM}_{10}:\mathrm{TE}_{10}$} case, reaches a $20\;\%$ error rate at $ka = 0.8$ and grows with increasing electrical size.
		
		\begin{figure}[t]
			\begin{center}
				\includegraphics[width=8.9cm]{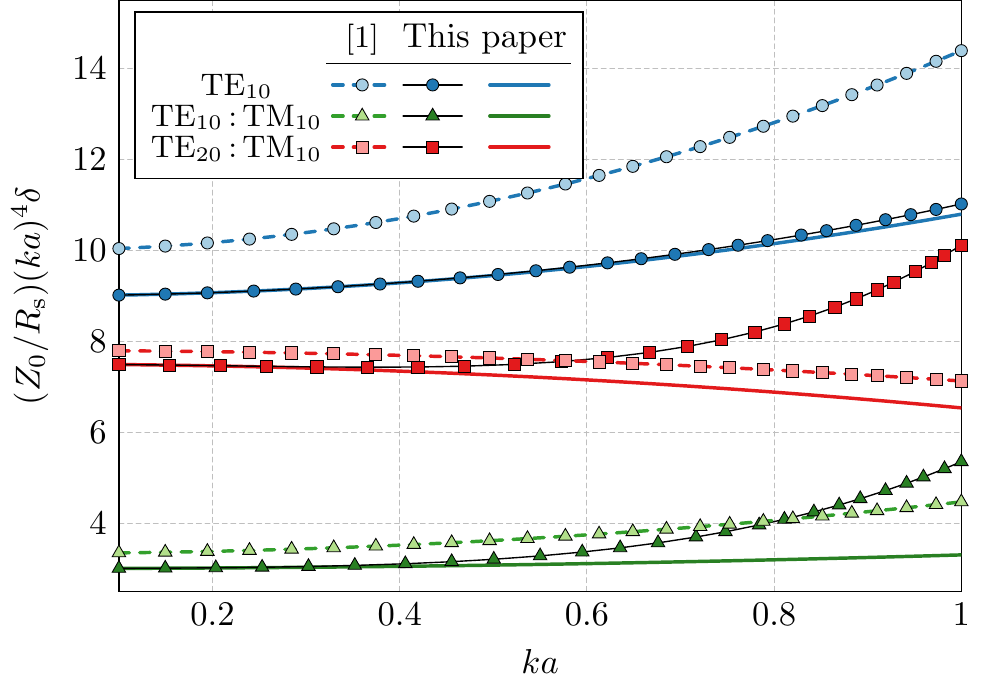}
				\caption{Comparison of results (14), (16) and (19) from~\cite{Pfeiffer_FundamentalEfficiencyLimtisForESA} with corresponding results of this paper. A comparison of the asymptotic (solid line) and full-wave (solid line with marks) expressions derived in this paper is also shown. The results correspond to a single spherical layer. The dissipation factors originating from~\cite{Pfeiffer_FundamentalEfficiencyLimtisForESA} were multiplied by a factor of two, since~\cite{Pfeiffer_FundamentalEfficiencyLimtisForESA} originally assumed two infinitesimally spaced resistive layers.}
				\label{fig1}
			\end{center}
		\end{figure}
		
		\begin{table}[t]
			\centering
			\caption{Comparison of asymptotic formulas for dissipation factor~$\delta$ normalized by~$\ZVAC / \Zsurf$ resulting from~\cite{Pfeiffer_FundamentalEfficiencyLimtisForESA} and from this paper. The results correspond to a single spherical layer.}
			\label{tab1}
			\begin{tabular}{ccc}
				$\left(\ZVAC / \Zsurf \right) \delta$ & Paper~\cite{Pfeiffer_FundamentalEfficiencyLimtisForESA} multiplied by 2 & This paper \\ \toprule
				TE$_{10}$ & $\displaystyle \frac{10}{\left(ka\right)^4} + \frac{22}{5\left(ka\right)^2}$ & $\displaystyle \frac{9}{\left(ka\right)^4} + \frac{9}{5\left(ka\right)^2}$ \\ \midrule
				TM$_{10}$\,:\,TE$_{10}$ & $\displaystyle \frac{10}{3\left(ka\right)^4} + \frac{34}{30\left(ka\right)^2}$  & $\displaystyle \frac{3}{\left(ka\right)^4} + \frac{3}{10\left(ka\right)^2}$  \\ \midrule
				TM$_{10}$\,:\,TE$_{20}$ & $\displaystyle\frac{78}{10 \left(ka\right)^4} - \frac{94}{140\left(ka\right)^2}$  & $\displaystyle \frac{15}{2 \left(ka\right)^4} - \frac{27}{28\left(ka\right)^2}$  \\ \bottomrule
			\end{tabular}
		\end{table}
		
		\subsection{A Note on Non-Resonant Current Distributions}
		
		The dissipation factors presented in Fig.~\ref{fig1} assume resonant current distributions due to the major dissipation cost of resonance tuning~\cite{Jelinek+etal2017}. Nevertheless, the non-resonant dissipation factors~\eqref{modalDelta1},~\eqref{modalDelta2} are also of importance. As an example, these analytical results can be used to validate more general dissipation bounds, such as those presented in~\cite{2016_Shahpari_Arxiv}. In particular, for a single spherical surface the results shown in~\cite[Eq.~18, version~5]{2016_Shahpari_Arxiv} suggest $\left(\ZVAC / \Zsurf \right) \delta = 6 / \left(2ka\right)^{2}$, while the first-order asymptotic expansion of \eqref{modalDelta2} gives $\left(\ZVAC / \Zsurf \right) \delta = 9 / \left(2 ka\right)^{2}$ for the lowest TM mode. The bound presented in~\cite{2016_Shahpari_Arxiv} is thus rather conservative for a spherical shell. It is also important to notice that, for small electrical sizes, non-resonant \mbox{electric-dipole-like} dissipation factors scale as $1 / \left(ka\right)^{2}$, while resonant dissipation factors scale as $1 / \left(ka\right)^{4}$, see~\cite{Jelinek+etal2017} for a more general exposition of this phenomenon.
		
		\section{Dissipation factor of two spherical layers}
		\label{Sec:TwoShells}
		
		The reduction of the dissipation factor by the specific composition of two resistive layers evokes the question of the general behavior of this setup. Specifically, assume that when forming a resonant current distribution~(\ref{modeCombination}), its constituents are yet another combination of spherical modes on two distinct layers of radius $a$ and radius $b<a$. The capacitive current will be formed as 
		\BE
		\label{modeCombination2}
		\Jcap = \Jcap_a + \betacap \Jcap_b
		\EE
		and the inductive current will be formed as
		\BE
		\label{modeCombination3}
		\Jind = \Jind_a + \betaind \Jind_b,
		\EE
		where it is assumed that currents of the same type (capacitive or inductive) are always formed by the same spherical mode. On the contrary, currents $\Jcap$ and $\Jind$ are always formed by two distinct spherical modes and are thus orthogonal with respect to complex power as well as lost power. Therefore, formula~\eqref{dissipGeneral} also remains valid in this case.
		
		Dissipation factors and normalized reactances corresponding to~\eqref{modeCombination2} and~\eqref{modeCombination3} read
		\begin{align}
			\label{Eq:deltaTECombined1}
			\delta_{\circledcirc,n}^\mathrm{TE} = & \frac{ \displaystyle {  A B + {\left| {\beta^{\mathrm{TE}}} \right|}^2 \frac{{ A }}{{ B}} } } {{ \displaystyle { A B + 2\mathrm{Re} \left[ {\beta^{\mathrm{TE}}} \right] + \frac{{\left| {\beta^{\mathrm{TE}}} \right|}^2}{A B}}}} \, \delta_{\circ,n}^\mathrm{TE}, \\
			\label{Eq:deltaTMCombined2}
			\displaystyle \delta_{\circledcirc,n}^\mathrm{TM} = & \frac{ \displaystyle { A B + {\left| {\beta^{\mathrm{TM}}} \right|}^2 \frac{B}{A} }} {{\displaystyle { A B + 2\mathrm{Re} \left[ {\beta^{\mathrm{TM}}} \right] + \frac{{{\left| {\beta^{\mathrm{TM}}} \right|}^2}}{{ A B}}}}}\,\delta_{\circ,n}^\mathrm{TM}, \\
			\label{Eq:lambdaTECombined3}
			\displaystyle \lambda_{\circledcirc,n}^\mathrm{TE} =& \frac{{ \displaystyle { A B + 2\mathrm{Re} \left[ {\beta^{\mathrm{TE}}} \right] + \frac{{{\left| {\beta^{\mathrm{TE}}} \right|}^2}}{{C B}}}}} {{ \displaystyle { A B + 2\mathrm{Re} \left[ {\beta^{\mathrm{TE}}} \right] + \frac{{{\left| {\beta^{\mathrm{TE}}} \right|}^2}}{{ A B}}}}} \, \lambda_{\circ,n}^\mathrm{TE}, \\
			\label{Eq:lambdaTMCombined4}
			\displaystyle \lambda_{\circledcirc,n}^\mathrm{TM} =& \frac{{ \displaystyle { A B + 2\mathrm{Re} \left[ {\beta^{\mathrm{TM}}} \right] + \frac{{{\left| {\beta^{\mathrm{TM}}} \right|}^2}}{{ A D }}}}} {{ \displaystyle { A B + 2\mathrm{Re} \left[ {\beta^{\mathrm{TM}}} \right] + \frac{{{\left| {\beta^{\mathrm{TM}}} \right|}^2}}{{ A B}}}}} \, \lambda_{\circ,n}^\mathrm{TM},
		\end{align}
		where
		\begin{eqnarray}
			\nonumber
			A &= \displaystyle\frac{\psi_n \left( ka \right)}{\psi_n \left( kb \right)}, \quad B & = \frac{\psi'_n \left( ka \right)}{\psi'_n \left( kb \right)}, \\
			C &= \displaystyle\frac{\chi_n \left( ka \right)}{\chi_n \left( kb \right)}, \quad D & = \frac{\chi'_n \left( ka \right)}{\chi'_n \left( kb \right)},
		\end{eqnarray}
		and where the $\circledcirc$ symbol denotes quantities corresponding to two spherical layers.
		
		As an example, the results of~\eqref{Eq:deltaTECombined1} for a $\mathrm{TE}_{10}$ mode are depicted in Fig.~\ref{fig3} and Fig.~\ref{fig4}.
		\begin{figure}[t]
			\begin{center}
				\includegraphics[width=8.9cm]{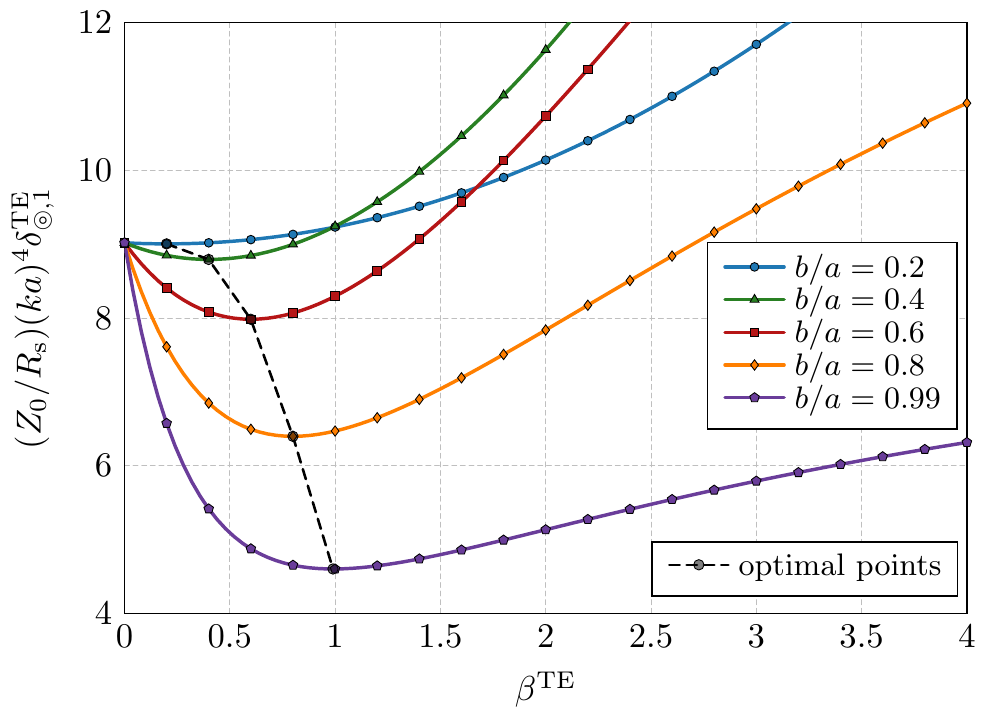}
				\caption{Normalized dissipation factor corresponding to a $\mathrm{TE}_{10}$ mode distributed on two spherical layers of radius $a$ and radius $b<a$. The results correspond to electrical size $ka=0.1$. A curve showing the minima of the dissipation factors is also shown.}
				\label{fig3}
			\end{center}
		\end{figure}
		A comparison of the curves in Fig.~\ref{fig3} and the curves in Fig.~\ref{fig1} shows that irrespective of ratio $b/a$, the $\mathrm{TE}_{10}$ current distribution on two spherical layers always results (for a specific $\beta^{\mathrm{TE}}$) in a lower dissipation factor than that of a single spherical layer\footnote{Notice that two infinitesimally spaced spherical layers assumed in \cite{Pfeiffer_FundamentalEfficiencyLimtisForESA} correspond to $A=B=C=D={\beta^{\mathrm{TE/TM}}}=1$ and thus exactly to two times lower dissipation factors in comparison to a single layer scenario.}. The optimal values of $\beta^{\mathrm{TE}}$ and $\beta^{\mathrm{TM}}$ are solutions to
		\begin{align}
			\label{Eq:betaOptTE}
			\left( \beta _\mathrm{opt}^\mathrm{TE} \right)^2 + \beta _\mathrm{opt}^\mathrm{TE} \frac{B}{A} \left(A^2-1\right) - B^2 & = 0, \\
			\label{Eq:betaOptTM}
			\left( \beta _\mathrm{opt}^\mathrm{TM} \right)^2 + \beta _\mathrm{opt}^\mathrm{TM} \frac{A}{B} \left(B^2-1\right) - A^2 & = 0.
		\end{align}
		The frequency sweep corresponding to the same scenario in Fig.~\ref{fig3}, but with optimal $\beta^{\mathrm{TE}}$, is shown in Fig.~\ref{fig4}. It can be observed that the reduction of dissipation factor for the two-layer scenario is almost independent of electrical size.
		
		\begin{figure}[t]
			\begin{center}
				\includegraphics[width=8.9cm]{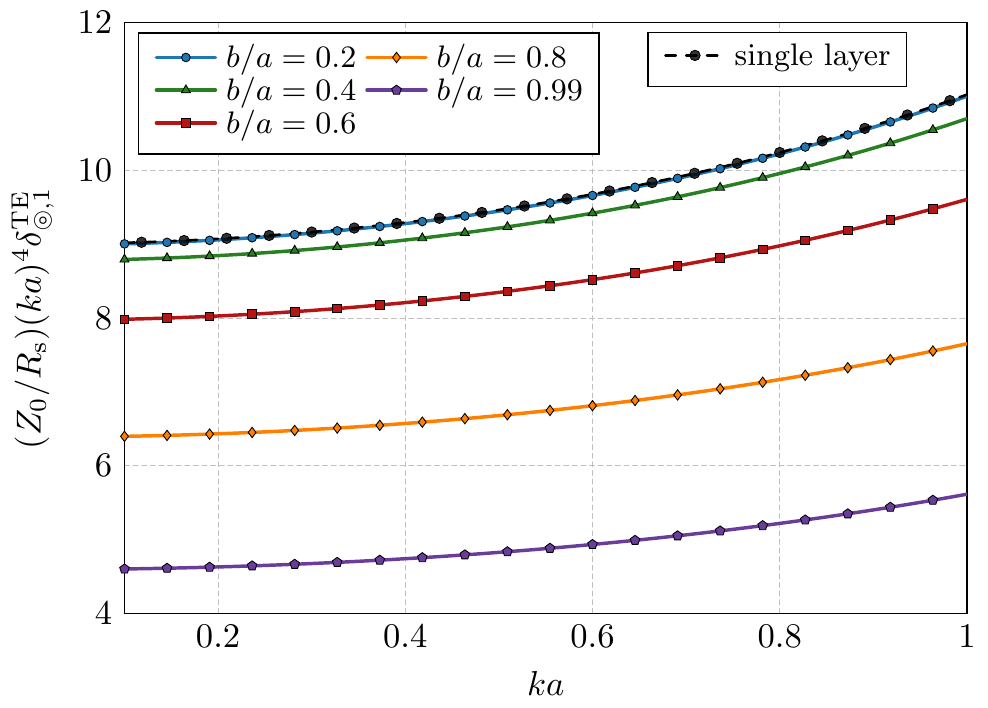}
				\caption{Normalized dissipation factor corresponding to a $\mathrm{TE}_{10}$ mode distributed on two spherical layers of radius $a$ and radius $b<a$. Optimal values of $\beta^{\mathrm{TE}}$ were used.}
				\label{fig4}
			\end{center}
		\end{figure}
		
		The attention is now turned to the lowest dissipation factor for the two-layer scenario. Drawing an analogy with Section~\ref{Sec:OneShell}, the lowest dissipation factor is assumed to be formed by a resonant combination of $\mathrm{TE}_{10}$ and $\mathrm{TM}_{10}$ modes. When composing this resonant combination, according to~(\ref{modeCombination}) in the two layer scenario, a first thought could be to set~$\beta^{\mathrm{TE}}$ and $\beta^{\mathrm{TM}}$ to their optimal values according to~(\ref{Eq:betaOptTE}) and (\ref{Eq:betaOptTM}), then form a resonant combination. This is, however, not an optimal choice as is shown in Fig.~\ref{fig5}. In the two-layer scenario, the normalized reactances $\lambda^{\CAP/\IND}$ are also functions of $\beta^{\mathrm{TE}}$ and $\beta^{\mathrm{TM}}$ making the minimum of the total dissipation factor an optimization problem with two variables. Depending on electrical size $ka$ and ratio $b/a$ the optimal values can deviate significantly from those predicted by~(\ref{Eq:betaOptTE}) and (\ref{Eq:betaOptTM}) for stand alone TM and TE modes, see Fig.~\ref{fig5}.
		
		\begin{figure}[t]
			\begin{center}
				\includegraphics[width=8.9cm]{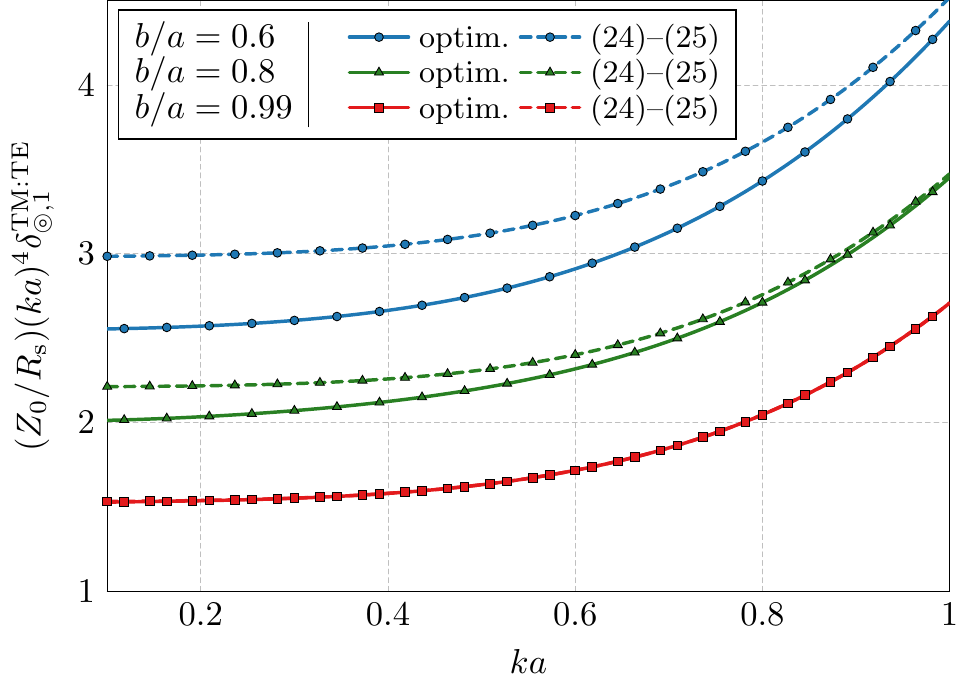}
				\caption{Normalized dissipation factor corresponding to a \mbox{$\mathrm{TM}_{10}:\mathrm{TE}_{10}$} combination distributed on two spherical layers of radius $a$ and radius $b<a$. Optimal values of $\beta^{\mathrm{TE}}$, $\beta^{\mathrm{TM}}$ were either evaluated according to (\ref{Eq:betaOptTE}) and (\ref{Eq:betaOptTM}) or by two variable optimization.}
				\label{fig5}
			\end{center}
		\end{figure}
		
		The optimal resonant combination \mbox{$\mathrm{TM}_{10}:\mathrm{TE}_{10}$}, shown in Fig.~\ref{fig5}, was proposed in \cite{Pfeiffer_FundamentalEfficiencyLimtisForESA} as a current density with the lowest dissipation factor from all free-space current distributions. An inductive extension of the analysis shown in this section, however, suggests that the addition of more layers should reduce dissipation even further.
		
		\section{Dissipation factor of multiple layers}
		\label{Sec:NShells}
		
		The case of more than two spherical layers is a straightforward extension of (\ref{Eq:deltaTECombined1})--(\ref{Eq:lambdaTMCombined4}). The number of terms in complex power~(\ref{Pcomplex}), however, increases and explicit relations become too long. It is also important to realize that the optimization of coupling parameters $\beta$ will attain more dimensions. Last, but not least, it is important to realize that we did not prove that the resonant combination of \mbox{$\mathrm{TM}_{10}:\mathrm{TE}_{10}$} modes on multiple spherical layers is the global minimizer to the resonant dissipation factor within spherical geometry. Due to the preceding reasons, this Section will compare a purely numerical approach with the analytical treatment.
		
		\begin{figure}[t]
			\begin{center}
				\includegraphics[width=8.9cm]{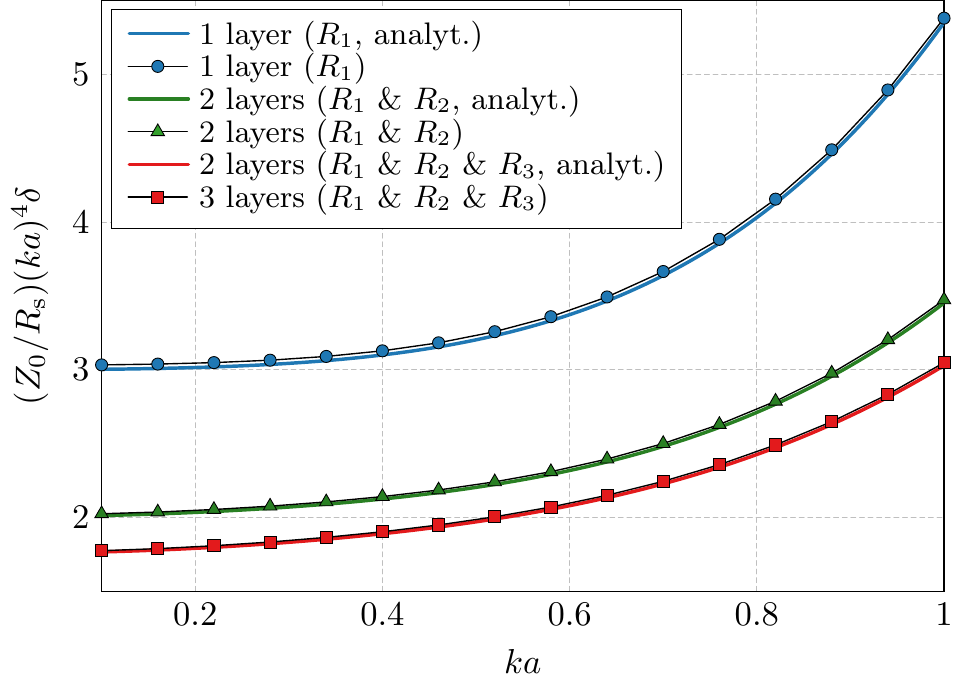}
				\caption{Normalized dissipation factors of the optimal self-resonant current densities distributed on one, two and three spherical layers of the same surface resistance. The radii of the layers $R_1=a$, $R_2=0.8a$, and $R_3=0.6a$ have been used. The analytical data correspond to the resonant \mbox{$\mathrm{TM}_{10}:\mathrm{TE}_{10}$} combination.}
				\label{fig2}
			\end{center}
		\end{figure}
		
		The numerical method used here, and described in~\cite{JelinekCapek_OptimalCurrentsOnArbitrarilyShapedSurfaces,CapekGustafssonSchab_MinimizationOfAntennaQualityFactor,2017_TEAT_Gustafsson_Tradeoff_Efficiency_Q}, is able to find the global minimizer for an arbitrary surface current support. The results for one, two, and three spherical shells are shown in Fig.~\ref{fig2} and compared to the analytical resonant combination of the \mbox{$\mathrm{TM}_{10}:\mathrm{TE}_{10}$} modes. Good agreement of the numerical and analytical results in Fig.~\ref{fig2} can be observed. A slight discrepancy can be attributed to the problem of comparing data corresponding to a perfect spherical surface with its triangularized (570 triangles per layer) counterpart. This allows us to finish this communication with the following statements:
		
		\begin{itemize}
			\item The addition of more spherical layers systematically reduces the dissipation factor, although with significantly diminishing returns;
			\item The resonant combination of \mbox{$\mathrm{TM}_{10}:\mathrm{TE}_{10}$} modes seems to give the lowest dissipation factor from all resonant current distributions, even in the multilayer scenario;
			\item The hypothesis from~\cite{Pfeiffer_FundamentalEfficiencyLimtisForESA} that the bound on the tuned dissipation factor is presented by a resonant combination of $\mathrm{TM}_{10}$ and $\mathrm{TE}_{10}$ spherical currents distributed on a single spherical surface is not valid.
		\end{itemize}
		
		It is worth noting that the last point is strongly connected to the optimization task addressed in~\cite{2013_Karlsson_PIER} and~\cite{Fujita_max_gain_IEICE_TE_2015} in which it is shown that a volumetric current density with the angular distribution of the dominant spherical mode and radial dependence of the spherical Bessel function exhibits a lower dissipation factor than the purely surface current distribution of the same angular dependence.
		
		\section{Conclusion}
		\label{Concl}
		
		Minimum dissipation factors corresponding to current densities distributed on multiple spherical layers have been found in an analytic or semi-analytic manner and have been proven to be valid by using a full-wave numerical method. Results corresponding to one and two spherical layers were also compared with existing works.
		
		It has been demonstrated that spherical modes can always be distributed on two spherical layers so as to lead to a smaller dissipation factor than that offered by a single spherical layer. This holds irrespective of electrical size or the ratio of the layer radii and does not depend whether a non-resonant or resonant combination of modes is formed. Moreover, the addition of more layers reduces the dissipation factor even further which indicates that a volumetric current density should be optimized in order to obtain a bound on dissipation factor. Since, however, radial currents are ineffective in producing radiation, the collection of separated spherical layers will lead to a solution close to the volumetric bound. Consequently, for a realistic antenna operating in free space environments, the surface currents can be considered as an approximate bound. However, for an antenna radiating in the presence of volumetric material objects, the volumetric current densities should be taken into account, since the surface current bound could be too pessimistic.
		


	\end{document}